\documentclass[aps,prl,twocolumn,amsmath,amssymb,superscriptaddress,floatfix,showpacs]{revtex4}

\usepackage{graphicx}
\usepackage{color}
\definecolor{DarkBlue}{rgb}{0.1,0.1,0.5}
\definecolor{Red}{rgb}{0.9,0.0,0.1}

\begin{document}
\title{Ising phases of Heisenberg ladders in a magnetic field}

\author{Karlo Penc}
\affiliation{
Research Institute  for  Theoretical Solid State  Physics   and
Optics, H-1525 Budapest, P.O.B.  49, Hungary}

\author{Jean-Baptiste Fouet}
\affiliation{
Institut Romand de Recherche Num\'erique en Physique des Mat\'eriaux (IRRMA), CH-1015 Lausanne}

\author{Shin Miyahara}
\affiliation{ 
Department of Physics and Mathematics, Aoyama Gakuin University, Sagamihara 229-8558, Japan}

\author{Oleg Tchernyshyov}
\affiliation{ 
Department of Physics and Astronomy, Johns Hopkins University, Baltimore, Maryland 21218, USA
}

\author{Fr\'ed\'eric Mila}
\affiliation{ 
Institute of Theoretical Physics, Ecole Polytechnique F\'ed\'erale de Lausanne, CH-1015 Lausanne, Switzerland
}

\date{\today}

\begin{abstract}

We show that Dzyaloshinskii--Moriya (DM) interactions can substantially 
modify the phase diagram of spin-1/2 Heisenberg ladders in a magnetic
field provided they compete with exchange. For non frustrated ladders, they 
induce a local magnetization along the DM vector that turns 
the gapless intermediate phase into an Ising phase with broken
translational symmetry, while for frustrated ladders, they extend the 
Ising order of the half-integer plateau to the surrounding gapless phases of the
purely Heisenberg case. Implications
for experimental ladder and dimer systems are discussed.  
\end{abstract}


\maketitle 

Spin ladders \cite{dagotto} have played an important role
in the field of strongly correlated systems over the past fifteen
years. They are among the best studied spin-gap systems, thanks to the
numerous experimental realizations in transition metal oxides
\cite{Dagotto99} and organometallic compounds \cite{Watson},
their intermediate phase in a magnetic field is one of the simplest
realizations of a Luttinger liquid\cite{Giamarchi}, and frustrating them
with diagonal rungs opens a magnetization plateau
at half-saturation\cite{mila}, a simple example 
of non-integer magnetization plateaux\cite{Oshikawa97}.

All this is true for purely SU(2) Heisenberg interactions, but 
in actual systems, anisotropic terms such as the Dzyaloshinskii-Moriya (DM)
interaction are often present. In ordered antiferromagnets, 
they open a small gap in the spin-wave spectrum and may induce
a small canting, but otherwise they can be neglected for most purposes. 
One may thus be tempted to conjecture that, for gapped
antiferromagnets, their effect will be negligible except in the gapless
phase induced by a magnetic field, where they will open a
gap. The intriguing properties of the famous  plateau
system SrCu$_2$(BO$_3$)$_2$\cite{miyahara_review}, in particular
the recent report that the translational symmetry remains broken above the
1/8 plateau\cite{Takigawareport}, suggest this might not be so simple
however. Indeed, significant DM interactions have been identified in this system, 
and the role they may play between the plateaux remains to be investigated.
In the same spirit, the nature of the phase transition of the intermediate phase of 
the spin gap system Cu$_2$(C$_5$H$_{12}$N$_2$)$_2$Cl$_4$\cite{chaboussant} 
is still an open issue, and the recent identification of
DM interactions in that system\cite{clemancey} opens new perspectives that
remain to be explored.

\begin{figure}[tb]
  \centering
  \includegraphics[width=8.0truecm]{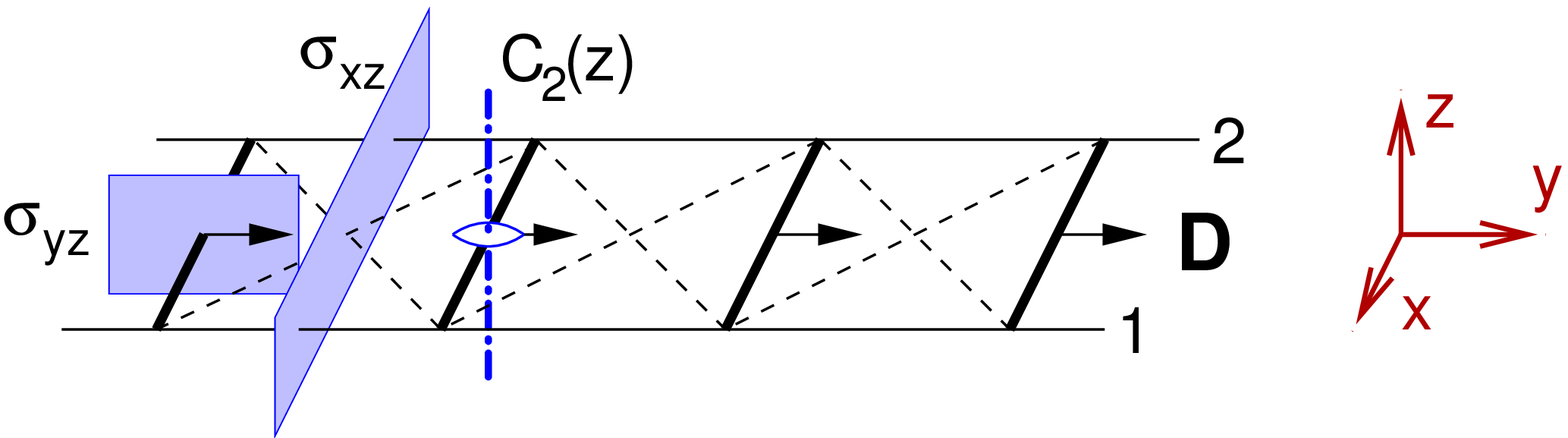}
  \caption{Sketch of the frustrated ladder of Eqs.~(\ref{eq:H-rung})-(\ref{eq:H-int}). The thick solid lines stand for $J$, the thin
ones for $J_\parallel$, the dashed ones for $J_\mathrm{X}$, and the arrows for $\mathbf D$. 
We have also shown the 3 generators of the space group: the $\sigma_{xz}$ and $\sigma_{yz}$ mirror planes and a twofold rotation axis $C_2(z)$.
   \label{ladder}}
\end{figure}

In this Letter, we report on an extensive investigation of the effect of DM interactions
in gapped systems in the context of spin ladders, and we show that
even very small DM interactions can induce Ising phase 
transitions with broken translational symmetry in all spin ladders, frustrated
or not, under appropriate conditions. To cover all aspects of the problem, we 
concentrate on a fairly general 
model  depicted in Fig. \ref{ladder}.  Its
Hamiltonian can be split into single-rung exchange, DM, and
Zeeman energies,
\begin{eqnarray}
 \mathcal{H_{\rm rung}} &=& 
 J \sum_{j} {\bf S}_{j,1}\cdot {\bf S}_{j,2} 
 + \sum_{j} {\bf D}\cdot ({\bf S}_{j,1} \times {\bf S}_{j,2}) \nonumber\\
 &&- \sum_{j} \mathbf H \cdot (\mathbf S_{j,1} + \mathbf S_{j,2}) 
\label{eq:H-rung}
\end{eqnarray}
and Heisenberg interactions between adjacent rungs,
\begin{eqnarray}
 \mathcal{H}_{\rm int} &=& 
 J_{\|}  \sum_{j} ({\bf S}_{j,1}\cdot {\bf S}_{j+1,1}+{\bf S}_{j,2}\cdot {\bf S}_{j+1,2})\nonumber\\
 &&+ J_\mathrm{X} \sum_{j} ({\bf S}_{j,1}\cdot {\bf S}_{j+1,2}+{\bf S}_{j,2} \cdot {\bf S}_{j+1,1})
\label{eq:H-int}
\end{eqnarray}
In this work, the magnetic field $\mathbf H=(0, 0, H)$ is applied
perpendicular to the DM vector $\mathbf D = (0, D, 0)$, which is taken to 
be uniform, the relevant configuration for ladders that are translationally
invariant but have no inversion center in the middles of the rungs. 
The full symmetry group of the system can be generated from three 
symmetry operations: $T\sigma_{yz}$, $C_2(z)$ and $T\sigma_{xz}$, 
as shown in Fig.~\ref{ladder}, where the additional time-reversal operator 
$T$ is needed to compensate for the presence of the magnetic field.  
The symmetry operations act both in spin and real space. 
  
The physics of this model is controlled by: i) The level of frustration 
of the Heisenberg exchange. Indeed, when $D=0$, a magnetization plateau at
half the saturation has been shown to open in strongly frustrated 
ladders ($1/3 < J_\mathrm{X}/J_{\|} < 3$), corresponding to a gapped
phase with a broken translational symmetry\cite{mila}. By contrast, non frustrated
ladders ($J_\mathrm{X}=0$ or $J_{\|}=0$) and weakly frustrated ladders
($0< J_\mathrm{X}/J_{\|} < 1/3$ or $0<J_{\|}/ J_\mathrm{X} < 1/3$)
remain in a gapless Luttinger liquid phase from zero magnetization to full
polarization. ii) The possible competition between exchange and DM interactions\cite{sato}.
In the present context, this can be understood as follows. Due to the DM interaction,
an isolated rung in a field develops opposite moments along $\mathbf D \times \mathbf H$,
on the two sites of the dimer\cite{miyahara}. If such dimers
are coupled in a ladder geometry, an antiferromagnetic inter-rung coupling $J_{\|}$ 
will compete with the DM interaction because $J_{\|}$ 
tends to align  the moments antiparallel along a leg. By contrast, an antiferromagnetic 
inter-rung coupling $J_\mathrm{X}$ will not compete since both $J_\mathrm{X}$ and
the DM interactions induce antiparallel moments along the diagonals.

The non-competing case ($J_\mathrm{X}/J_{\|}>1$) has
been studied previously\cite{fouet}. The DM interaction opens a gap in the gapless
incommensurate phase, as expected, but only induces a small shift of the Ising plateau transition
in strongly frustrated ladders. In the following, we will concentrate on the competing case  ($J_\mathrm{X}/J_{\|}<1$).

The simplest way to map out the phase diagram is to use a Hartree 
variational function of the form 
\begin{equation}
\psi = \prod_j \left( \cos \frac{\theta_j}{2}  |T\rangle + 
e^{i\varphi_j} \sin \frac{\theta_j}{2} |S\rangle\right)
\end{equation}
with $|T\rangle = |\uparrow_{1} \uparrow_{2}  \rangle$ and 
$|S\rangle = (|\downarrow_{1} \uparrow_{2}\rangle - |\uparrow_{1} \downarrow_{2} \rangle)/\sqrt{2}$ 
on the $j$-th rung \cite{momoi}.  The trial state is characterized by the symmetric and 
antisymmetric magnetizations of a rung, 
$\mathbf m_{j\pm}=\langle \mathbf S_{j,1} \pm  \mathbf S_{j,2}\rangle$.
The nonzero components are
\begin{equation}
m^x_{j,-} + i m^y_{j,-} = \sin \theta_j e^{i \varphi_j} / \sqrt{2}, \ 
m^z_{j,+} = \cos^2 (\theta_j/2).
\label{eq:magns}
\end{equation}

\begin{figure}[tb]
  \centering
  \includegraphics[width=8.5truecm]{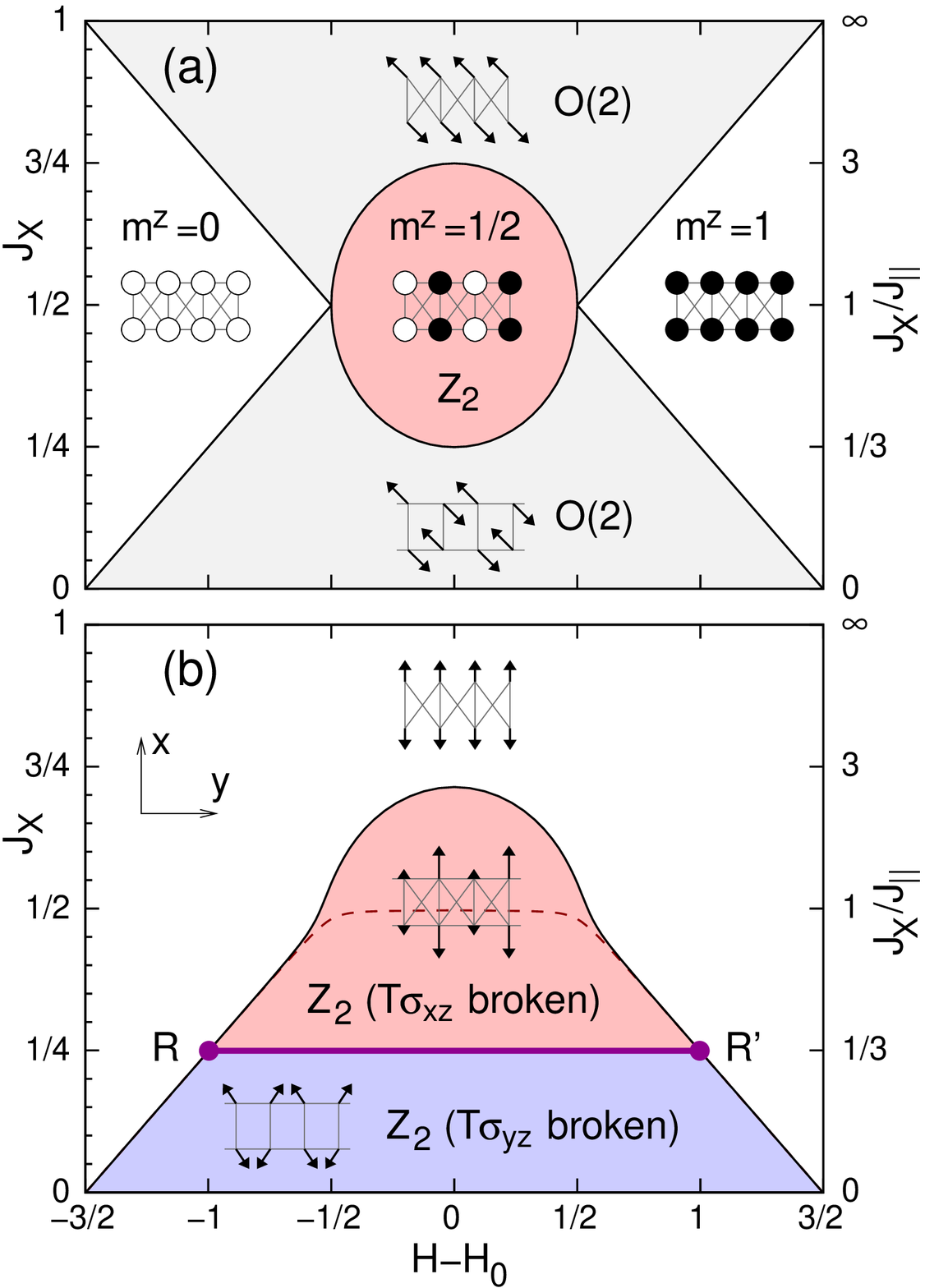}
  \caption{(Color online) Variational phase diagram for (a) $D=0$ and (b) $D/(J_\mathrm{X}+J_{\parallel})=0.1$. On the left and bottom axes, $J_\mathrm{X}$ and $H-H_0$, where $H_0=J+ (J_\mathrm{X}+J_\|)/2$, are given in the units of $J_\mathrm{X}+J_\|$. In (a), the spins forming singlets and triplets in the gapped phase are represented by open and solid circles. In the gapless $O(2)$ phase the arrows shows a typical mean-field ground state. In (b), the $O(2)$ symmetry is broken by finite $D$: a unique state is selected in the non frustrated case, an a two-fold degenerate state in the frustrated case. The Ising phase of the plateau in (a) develops continuously by turning on $D$. Along the red dashed line the site--factorized wave function is an exact wave function in the large $J$ limit. R and R' denote the tricritical points.
   \label{phasediagram}}
\end{figure}
The variational phase diagram without DM interaction is shown in 
Fig.~\ref{phasediagram}(a). It is symmetric 
with respect to the exchange of $J_\mathrm{X}$ and $J_{\|}$ and contains all the phases found 
previously \cite{fouet}: 
two fully symmetric integer plateau states with $m^z_{j,+} = 0$ and 1;  
a half-integer plateau phase with a broken $Z_2$ translational symmetry, 
$m^z_{j,+} - 1/2 \propto (-1)^j $,
and a gapless phase with a spontaneously broken O(2) symmetry (the
latter being an artifact of the mean-field approximation).

The variational phase diagram in the presence of the DM interaction is
shown in Fig.~\ref{phasediagram}(b). It is clearly related to the
O(2)-symmetric case: some of the transition lines lie close to their
$D=0$ counterparts.  However, there are some notable differences.  In
the noncompeting regime, $J_\mathrm{X}/J_{\|} > 1$, the phase
transitions between the integer plateaux and the intermediate gapless
phase are replaced with crossovers, in agreement with previous work
\cite{fouet}.

Far more dramatic changes can be seen in the competing regime,
$J_\mathrm{X}/J_{\|} < 1$.  At strong frustration, $1/3 <
J_\mathrm{X}/J_{\|} < 1$, the Ising-ordered phase
has substantially expanded its boundaries by absorbing the
incommensurate phases.  A \textit{distinct} Ising order
appears for zero or weak frustration, $0\le J_\mathrm{X}/J_{\|} < 1/3$.  In
the upper Ising phase, the order parameter is the staggered component
of the magnetizations $m^z_{j,+}$ and $m^x_{j,-}$. It breaks
$T\sigma_{xz}$.  In the lower Ising phase,
the order parameter is the staggered component of $m^y_{j,-}$.
It breaks $T\sigma_{yz}$. Since $T\sigma_{yz} C_2(z) T\sigma_{xz}$ is the translation 
operator, both phases break the translational invariance as well. In any of the two phases, 
only one generator of the space group is broken, thus the transitions from the uniform phase
into these phases are continuous. At the variational level, the
transition between the two Ising phases is first order, but numerical results and field-theory
arguments suggest that the transition is made continuous by quantum critical fluctuations (see below).

\begin{figure}[tb]
  \centering
  \includegraphics[width=8.5truecm,angle=0]{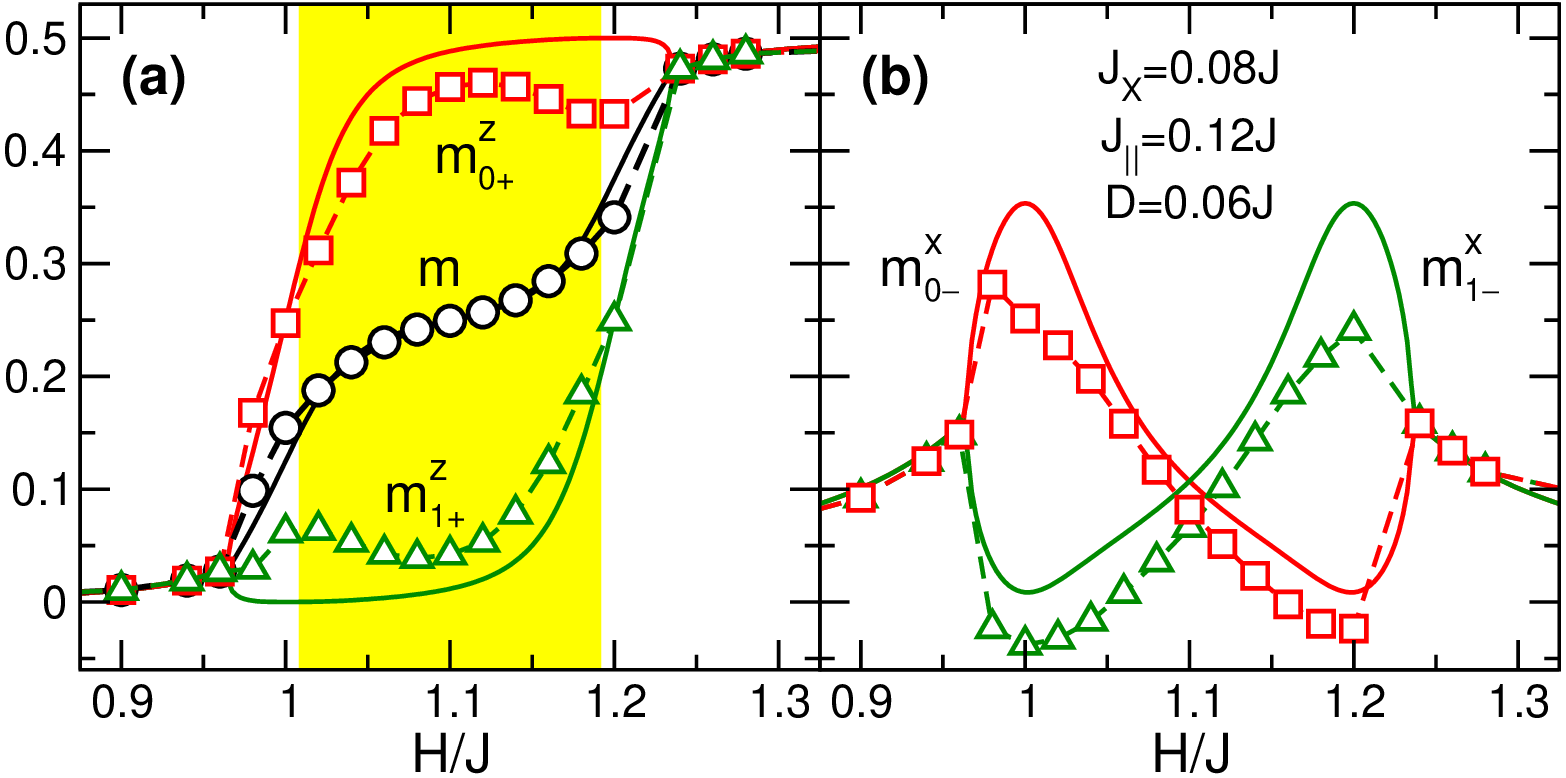}
  \caption{(Color online) (a) Uniform magnetizations of neighboring rungs ($m^z_{0+}$ and $m^z_{1+}$) and average ($m$) in the upper Ising phase and (b) staggered magnetizations $m^x_{0+}$ and $m^x_{1+}$. Solid lines: variational approach; symbols and dotted lines: DMRG for 58 sites; shaded area: plateau phase when $D=0$.
   \label{DMRG1}}
\end{figure}

\begin{figure}[tb]
  \centering
 \includegraphics[width=6truecm,angle=0]{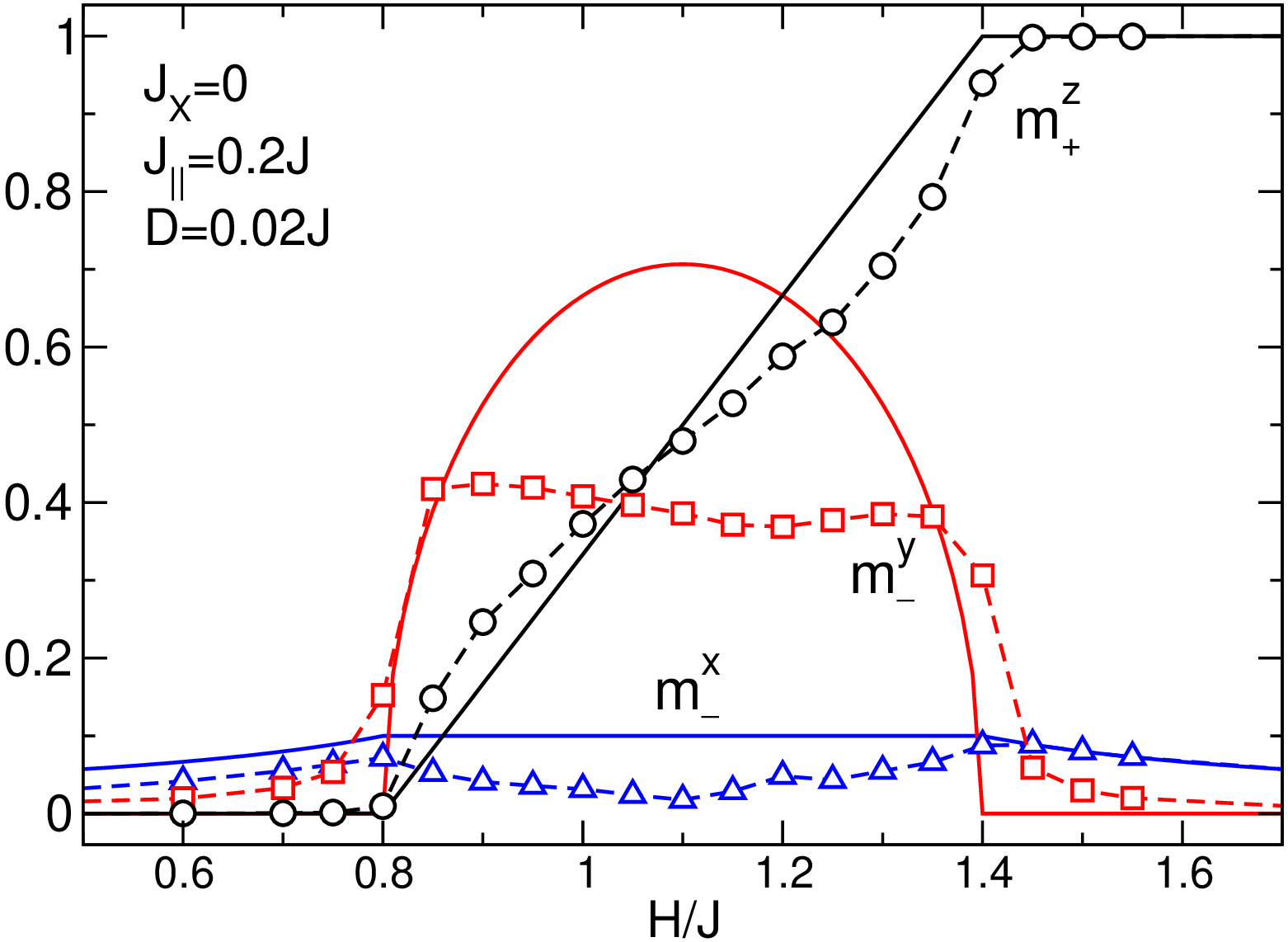}
  \caption{(Color online) Magnetizations in the lower Ising phase. Solid lines: variational approach; symbols and dotted lines: DMRG for 80 sites.
   \label{DMRG2}}
\end{figure}

To test the reliability of the variational approach, we have obtained the ground state 
and magnetization distributions for the ladder model (\ref{eq:H-rung}-\ref{eq:H-int}) 
using the Density Matrix Renormalization Group (DMRG)
for two representative cases: the unfrustrated
ladder ($J_\mathrm{X}=0$) and a strongly frustrated one ($J_\mathrm{X}/J_{\|}=2/3$), and
we have obtained 
strong evidence for the existence of the two Ising phases.
The results for relevant local magnetizations
are compared with the variational results in Figs.~\ref{DMRG1} and \ref{DMRG2}.  In the upper Ising phase 
($J_\mathrm{X}/J_{\|}=2/3$, Fig.~\ref{DMRG1}), the agreement is nearly perfect and the presence of 
a broken symmetry far outside the half-integer magnetization plateau is well established:
magnetizations $m^z_{j,+}$ of two neighboring rungs exhibit a finite difference
in almost the entire range of fields between the two integer magnetization plateaus. 
For $J_\mathrm{X}/J_{\|}=0$ (see Fig.~\ref{DMRG2}),
we have included a small staggered field to allow the finite system to maintain a 
non-zero staggered magnetization along the $D$ vector. 
We find a
qualitative agreement between the variational and numerical results.  Quantitatively, the
mean-field approach overestimates the order parameter by as much as a factor of 2, 
presumably the result of neglecting quantum fluctuations.  

\begin{figure}[tb]
  \centering
  \includegraphics[width=3.5truecm]{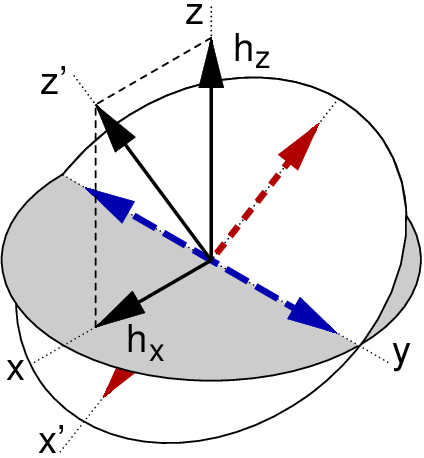}
  \caption{(Color online) 
Schematic representation of the phase transitions. 
Away from the $\Delta=1$
Heisenberg line, the $\mathcal{H}_{\rm eff}$ loses the axial symmetry 
around the total field: the spins 
develop staggered order along the long dashed blue arrows if $\Delta<1$ and along
the short dashed red arrows if $\Delta>1$.
   \label{ellipse}}
\end{figure}

To understand the physics of the ordered states and phase transitions
between them we turn to the strong-coupling limit where the rung
exchange $J$ of Eq.~(\ref{eq:H-rung}) dominates.  The low-energy physics
of the model then reduces to that of a spin-1/2 XXZ chain \cite{mila},
\begin{eqnarray}
\mathcal{H}_{\rm eff} & =&  j\sum_{i}
 (\sigma^x_i \sigma^x_{i+1}+\sigma^y_i \sigma^y_{i+1} 
 + \Delta \sigma^z_i \sigma^z_{i+1}) \nonumber\\
 & - & \sum_{i} (h_x \sigma^x_i + h_z \sigma^z_i),
  \label{eq:hameff}
\end{eqnarray}
with $j = J_{\|}-J_\mathrm{X}$ and $\Delta=(J_\mathrm{X}+J_{\|})/2(J_{\|}-J_\mathrm{X})$. 
The longitudinal field is related 
to the field in the original problem, $h_z = H - J - (J_\mathrm{X} + J_{\|})/2$, while the transverse 
field comes from the DM interactions, $h_x = D/\sqrt{2}$, as discussed previously \cite{oshikawa}. 
The spin-1/2 operators $\sigma$ are related to the original spin operators 
in the following way:
\begin{equation}
\sigma^\alpha_j=P \frac{S^\alpha_{j,1}-S^\alpha_{j,2}}{\sqrt2}P,\ \  
\sigma^z_j=P \left (S^z_{j,1}+S^z_{j,2}-\frac{1}{2}\right)P
\nonumber
\end{equation}
where $\alpha = x,y$ and $P=(S^+_{j,1}-S^+_{j,2})(S^-_{j,1}-S^-_{j,2})/2$ is the projector 
onto the subspace $\{|S\rangle , |T\rangle \}$. Nonzero averages 
$\langle \sigma^x_j \rangle$ and $\langle \sigma^y_j \rangle$ signal the appearance of 
antisymmetric magnetizations $m^x_{j,-}$ and $m^y_{j,-}$, while  
$\langle \sigma^z \rangle$ translates into the symmetric magnetization $m^z_{j,+}-1/2$, 
cf. Eq.~(\ref{eq:magns}).

The XXZ model in a tilted uniform field (\ref{eq:hameff}) has been
studied previously \cite{dmitriev,caux03}.  The magnetic order is
lost across the line of Ising transitions, $h_z \approx (\Delta+1)j$,
into the fully magnetized plateaus.  The two Ising phases are
separated by the Heisenberg line $\Delta = 1$ along which
the system has an axial symmetry.
For $h = |\mathbf h| < 2j$, the system is gapless and can be viewed as a Luttinger liquid of
magnons, whose dimensionless compressibility $K$ \cite{Giamarchi}
varies between 1 (a dilute magnon gas near the fully magnetized state)
and 1/2 (a dense magnon fluid at low magnetization) \cite{Haldane80}.

To discuss the physics away from the Heisenberg point, it is
convenient to make a global rotation in the $xz$ plane, defining a new
quantization axis $z'$ aligned with the total field $\mathbf h$.  When $\Delta \neq 1$, this
rotation generates new terms in the Hamiltonian.  The nature of the
phase transitions becomes particularly transparent in the vicinity of
the tricritical points R and R' located at $\Delta = 1$, $h =\pm 2j$,
where the system can be described as a dilute magnon gas.  Then the main effect is the lowering of the axial symmetry by a  weak anisotropy induced in the $x'y$ plane,
\begin{equation}
\lambda \sum_j (\sigma^{x'}_j \sigma^{x'}_{j+1} - \sigma^y_j \sigma^y_{j+1})
= \frac{\lambda}{2} \sum_j (\sigma^+_j \sigma^+_{j+1} + \sigma^-_j \sigma^-_{j+1}), 
\label{eq:easy-axis}
\end{equation}
where $\lambda \sim (\Delta-1){h_x}^2/8j$. The magnon
Luttinger liquid becomes gapped by developing a staggered magnetization along $y$ if $\Delta<1$ 
(breaking $T\sigma_{xy}$ \cite{generators}) 
or 
along $x'$ if $\Delta>1$ (breaking $T\sigma_{xz}$ \cite{generators}), see Fig.~\ref{ellipse}.  
The anisotropy (\ref{eq:easy-axis})
represents a pairing field for the magnons.  From the scaling
dimension of the pairing operator we infer that the energy gap in the
magnon spectrum should scale as $E_g \propto |\lambda|^\nu$ with a
nonuniversal exponent $\nu = K/(2K-1)$.

The global rotation also generates a 3-body term \cite{dmitriev}, 
\begin{equation}
\frac{(\Delta-1)h_x}{4}\sum_{j}(\sigma^+_j + \sigma^-_j)
(\sigma^-_{j-1} \sigma^+_{j-1} + \sigma^-_{j+1} \sigma^+_{j+1}).  
\end{equation}
which has only a minor effect on the
physics near the critical line separating the Ising phases.  While
low-energy magnons live near lattice momentum $\pi$, this term creates
1- and 3-magnon states with a total momentum 0, which are thus
high-energy excitations.  Therefore the 3-body term does not affect
the low-energy physics to the first order.  In the second order it
generates 2- and 4-magnon terms that simply renormalize the magnon
velocity, compressibility $K$, and pairing field $\lambda$ but do
not change the physics qualitatively.  The existence of two distinct
ordered phases is confirmed by an exact solution on a special line in
the phase diagram where the mean-field approach yields the correct
result \cite{dmitriev}.  The Heisenberg line $\Delta=1$ is guaranteed
to be critical by the O(2) symmetry.

To summarize, we have shown that even a weak DM interaction can
substantially alter the phase diagram of ladders,
provided that it competes with exchange.  For strongly frustrated ladders, 
the phase with a broken translational symmetry expands 
to include the incommensurate regions surrounding the 
fractional plateau, while for unfrustrated and weakly frustrated ladders,  
the gapless phase undergoes an Ising transition that also breaks translation, 
but with a different symmetry.  Beyond ladders, we expect to find similar
effects in higher dimensional models of coupled spin-1/2 dimers, whenever DM and exchange compete. In fact, the broken translational symmetry above 
the 1/8 plateau of SrCu$_2$(BO$_3$)$_2$ might be an example of the frustrated case, 
and the ordered intermediate phase of  Cu$_2$(C$_5$H$_{12}$N$_2$)$_2$Cl$_4$ 
of the unfrustrated one. However, taking into account the specific (and complex) 
geometries of these systems is necessary to go beyond these qualitative observations.

We  acknowledge enligthening discussions 
with C. Berthier, M. Horvatic, S. Matsubara, 
B. Normand and M. Takigawa.
This work was supported by the Hungarian OTKA Grant Nos. T049607 and K62280, 
by the Swiss National Fund, by MaNEP, and by the US National Science 
Foundation Grant No. DMR-0348679.

\end{document}